\documentclass[sigconf]{acmart}
\AtBeginDocument{%
  }

\copyrightyear{2025}
\acmYear{2025}
\setcopyright{acmlicensed}\acmConference[K-CAP '25]{Knowledge Capture Conference 2025}{December 10--12, 2025}{Dayton, OH, USA}
\acmBooktitle{Knowledge Capture Conference 2025 (K-CAP '25), December 10--12, 2025, Dayton, OH, USA}
\acmDOI{10.1145/3731443.3771344}
\acmISBN{979-8-4007-1867-0/2025/12}
\acmISBN{978-1-4503-XXXX-X/2018/06}

\usepackage{multirow}
\usepackage{textcomp}
\usepackage{url}
\usepackage{verbatim}
\def\dm{GUI Interaction}
\def\nlc{Textual Command}
\def\nl{Natural Language}
\begin{document}

\title{Natural Language Interaction for Editing Visual Knowledge Graphs}

\author{Reza Shahriari}
\orcid{0009-0003-0196-2199}
\affiliation{%
  \institution{University of Florida}
  \city{Gainesville}
  \state{Florida}
  \country{USA}
}
\email{rshahriari@ufl.edu}

\author{Eric D. Ragan}
\orcid{0000-0002-7192-3457}
\affiliation{%
  \institution{University of Florida}
  \city{Gainesville}
  \state{Florida}
  \country{USA}}
\email{eragan@ufl.edu}

\author{Jaime Ruiz}
\orcid{0000-0002-9139-6172}
\affiliation{%
  \institution{University of Florida}
  \city{Gainesville}
  \state{Florida}
  \country{USA}
}
\email{jaime.ruiz@ufl.edu}

\renewcommand{\shortauthors}{Shahriari et al.}

\begin{abstract}
  % \eric{short abstract}
% \TODO{Abstract should make it clear it's a VIS paper}
Knowledge graphs are often visualized using node-link diagrams that reveal relationships and structure.
In many applications using graphs, it is desirable to allow users to edit graphs to ensure data accuracy or provides updates.
Commonly in graph visualization, users can interact directly with the visual elements by clicking and typing updates to specific items through traditional interaction methods in the graphical user interface.
However, it can become tedious to make many updates due to the need to individually select and change numerous items in a graph.
Our research investigates natural language input as an alternative method for editing network graphs.
We present a user study comparing GUI graph editing with two natural language alternatives to contribute novel empirical data of the trade-offs of the different interaction methods.
The findings show natural language methods to be significantly more effective than traditional GUI interaction.

\end{abstract}





\maketitle

\begin{figure*}
\centering
\includegraphics[width=1.8\columnwidth]{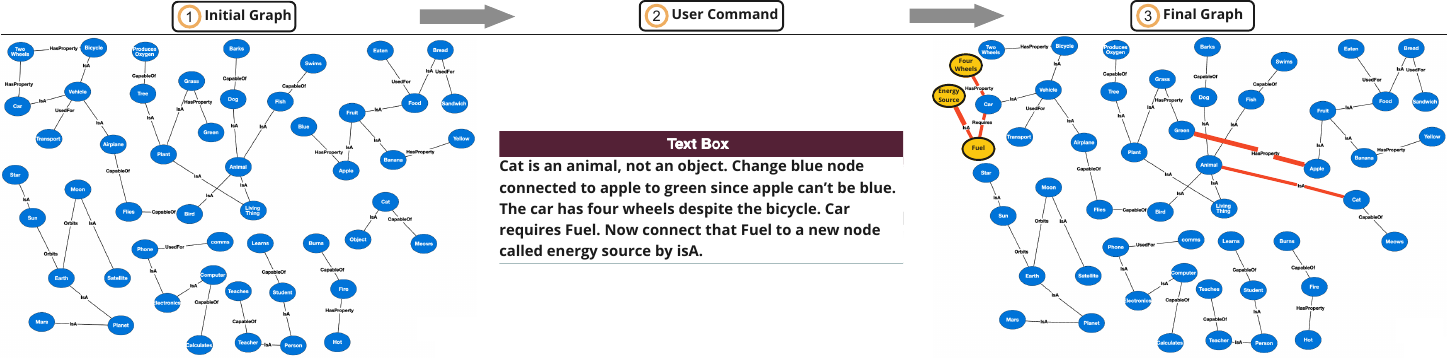} 
\caption{Screenshot of the interface visualizing relationships from the ConceptNet~\cite{speer2017conceptnet} dataset.}
\label{fig:vis}
\end{figure*}

\section{Introduction}
A knowledge graph organizes information by connecting entities and their attributes within a structured, semantic framework.
To support meaningful reasoning and interpretation, knowledge graphs are often visualized using node-link diagrams that reveal relationships and structure.
% For instance, in bioinformatics, knowledge graphs map complex biological data such as gene interactions and disease associations \cite{munarko2023casbert,theocharidis2009network,thattai2001intrinsic}.
% Other examples are found in web development, where graphs improve search engines by structuring data about entities, enhancing search accuracy and relevance \cite{heist2020knowledge,orlandi2018leveraging}.
% For many practical data applications, it is beneficial for network representations to capture more than only the binary presence or absence of connections between items;
% it is therefore common to augment graph data with more information with additional descriptive properties of objects or types of relationships between them.
Maintaining accurate data is essential for applications relying on network data, especially in dynamic environments where real-time data impacts outcomes~\cite{ramamritham1993real, ramamritham2004real}. 
Graph updates involve changes in items (nodes) or relationships (links), and changes to additional properties are also common for knowledge graphs.
% For example, in real-time databases, temporal consistency is critical for applications like air traffic control systems, where any delay or inaccuracy in data could lead to severe safety risks as flight path decisions depend on up-to-the-second information~\cite{ramamritham2004real}.
Regular interaction with these structures, such as editing, expanding, and correcting, ensures they remain reliable.
% It helps fix errors, add missing information, and connect related data, improving scalability and support for new features.
Editing graph relationships is especially useful in human-in-the-loop (HITL) applications, such as correcting or enriching data \cite{monarch2021human, fu2021hoops, bikaun2024cleangraph}.

While node-link graph visualizations are excellent for easy interpretation of small graphs~\cite{burch2021dynamic}, common forms of interaction for visual graph editing can be tedious and time-consuming for a large number of edits.
% Performing numerous interactions for simple changes in nodes or links can also be inefficient, leading to delays and increased effort.
Typically, interactions with graphs involve using visual interfaces where users select individual nodes or edges to make changes directly within the graph \cite{mcguffin2009interaction}.
This process includes actions like clicking nodes to view or edit properties, dragging to explore relationships, and drawing or deleting edges to modify connections. 
These methods tend to be easily understood and align well with the familiar interaction philosophy of \textit{direct manipulation}~\cite{shneiderman1983direct}.
However, direct GUI interactions in graphs also typically require execution of each individual action.
Consequently, due to the importance of connectivity in network data, basic updates often require multiple actions.
As graph size or complexity grows, GUI interactions become more cumbersome and time-consuming, increasing user effort.

To address the limitations of traditional GUI-based interaction, our research studies alternative methods for editing visual graph representations.
We investigate how natural-language processing (NLP) techniques can increase the efficiency of interacting with graphs using text-based inputs. 
We demonstrate interaction methods that allow users to describe multiple graph relationships and properties through language instead of a sequence of individual edit operations. 
To evaluate these methods, we compare natural language (NL) inputs with structured textual commands, which offer precision but less accessibility for non-technical users.
This comparison highlights trade-offs between flexibility and efficiency in graph editing tasks.
Also, we conducted an experiment comparing three interaction types: \dm, \nlc, and \nl.
The study involved user updates to labeled knowledge graphs with varied size and expected changes. 
Results show that natural language enables effective data manipulation, offering advantages over traditional command formats and supporting more accessible, efficient interaction.

% We conducted an experiment to assess the capability of text-based natural-language interaction for editing graphs. 
% The experiment compares three types of interactions for graph editing: \dm, \nlc, and \nl.
% To account for different types of edits, the study required forms of user updates of knowledge graphs with labeled nodes and edges, and the study design also varied size and expected amount of changes in each graph.
% This research demonstrates the feasibility and advantages of using natural language for data manipulation, offering a comparison to traditional command formats. 
% It highlights the potential for enhancing user interaction by allowing descriptive language, making data tasks more accessible and efficient. 
% These insights can influence the design of future data interaction systems.

\section{Background}
\subsection{Knowledge Graph Visualization}
Knowledge graphs illustrate the connections between entities, enabling an understanding of relationships.
Various visualization formats support interpretation, with node-link diagrams being the most common, using points (e.g., circles or boxes) for entities and lines for connections \cite{munzner2014visualization}.
Another common representation is adjacency matrix \cite{munzner2014visualization}, where nodes are arranged along the vertical and horizontal edges of a square, and connections are indicated by flagging the cell at the intersection of two nodes. 
% Additional visualization techniques offer different strengths.
% For instance, arc diagrams are an alternative representation where nodes are placed along a line, and edges are drawn as arcs above the line~\cite{wattenberg2002arc}. While less commonly used, arc diagrams can help highlight specific patterns like clustering and repetition within ordered data ~\cite{wattenberg2002arc}.
% These types of visualization are commonly used in various contexts, including social network analysis~\cite{blythe1995effect}, law enforcement~\cite{baccara2008organize, mcillwain1999organized, sparrow1991application}, to help understand complex relationships and structures within a dataset \cite{saket2014node}.
Moreover, multiple tools \cite{dunne2012graphtrail,henry2007nodetrix} combined these representations to allow a hybrid representation of network data that can take advantage of multiple representations.
For instance, NodeTrix~\cite{henry2007nodetrix} combines node-link diagrams and adjacency matrices into a single, cohesive representation to ease interpretation.
% The global structure of the network is displayed as a node-link diagram, while communities, which are often dense and difficult to interpret using only node-link representations, are shown as matrices.
% This hybrid approach allows for better readability and detailed analysis of community structures while still preserving the overall network's layout.
% Users can interactively switch between node-link and matrix representations.
Similarly, GraphTrail~\cite{dunne2012graphtrail} integrates various visualizations, such as bar charts, tag clouds, and node-link diagrams, to analyze large multivariate networks. 
By allowing users to pivot between different aggregates of nodes and edges based on their attributes, GraphTrail supports dynamic exploration of heterogeneous networks.
% Its strength lies in enabling users to transition seamlessly between visual representations, helping them uncover patterns that may not be easily visible in a single representation.

% Existing literature explored different types of visualization and interaction with these types of structures~\cite{herman2000graph, tominski2006fisheye, van2004interactive}.
% One of the interaction techniques is the \textit{zoom and pan} for navigating large graph structures, allowing users to explore content effectively without losing visual clarity \cite{herman2000graph}.
% While geometric zoom simply magnifies graph content, semantic zooming enhances the display by revealing additional details as users focus on specific areas.
% However, zooming can cause a loss of contextual information, which can hinder usability.
% To address this, \textit{focus-plus-context} techniques~\cite{baudisch2001focus,doleisch2003interactive} have been developed, enabling users to maintain a broader context while focusing on specific details, often complementing traditional \textit{zoom and pan} functionalities in complex data visualization systems \cite{herman2000graph}.

Different studies have applied various techniques \cite{tominski2006fisheye,van2004interactive} to achieve this balance.
For example, Tominski et al. \cite{tominski2006fisheye} presented fisheye tree views for exploring hierarchical trees as well composite lenses combined with several lens techniques to facilitate the exploration of local graph information.
They found that these tools provided enhance navigation of complex graph structures, with their potential in visualizing clustered graphs.
% Additionally, Van Ham and van Wijk \cite{van2004interactive} developed a visualization approach that combined semantical and geometrical distortions.
% The primary objective was to create scalable, interactive visualizations of small-world graphs known for their small diameters and high clustering.
% Their method was tested on a graph of 500 cross-referenced artists, demonstrating that the visualization effectively displayed the clustering structure of small-world graphs despite the challenges posed by their high connectivity.
% These visualization and interaction techniques highlight the importance of efficient user interfaces for exploring and manipulating complex network data. 
While past work improved navigation and understanding of large graphs, more effective methods are still needed for editing, especially in knowledge graphs with rich semantic information.
% Building on these, our research investigates how natural language processing can be integrated into visualization tools to facilitate graph editing.

\subsection{Editing Network Data}

After visualizing a network data structure, it is important to keep it current and accurate by incorporating new information and correcting any errors.
One approach is using \textit{direct manipulation}, that involves actions such as clicking, dragging, and dropping.
These interactions provide immediate feedback and enhance the user's sense of control \cite{shneiderman1983direct}. 
% These techniques focus on interaction through visible objects and actions, allowing users to perform operations that can be easily and quickly reversed. 
The goal is to create interfaces that minimize the need for complex command languages and intermediary steps \cite{shneiderman1983direct}.
Optimizing interaction through \textit{direct manipulation} required further exploration of various aspects to understand the reasons behind the usability or challenges that users encountered with these interfaces.
For example, Hutchins et al. \cite{hutchins1985direct} studied the cognitive aspects of direct manipulation to understand why these interfaces feel natural and engaging to users and to identify potential issues.
They found that reducing the cognitive distance between what users intend to do and how they interact with the system, along with providing immediate feedback, are key factors in the effectiveness of direct manipulation interfaces.
% However, they also pointed out that these interfaces can struggle with repetitive tasks, where symbolic descriptions or commands might be more efficient.

% Moreover, understanding the weaknesses of direct manipulation has led developing improved techniques and features \cite{cardelli1988building,kurtenbach1991issues}.
% For instance, Cardelli et al. \cite{cardelli1988building} identified the difficulties in creating polished user interfaces and developed a method to simplify the process by using a user interface editor for direct manipulation, separating the interface from the application. 
% The goal was to make building, modifying, and customizing user interfaces simpler for developers and accessible for users without technical knowledge.
% His approach allowed developers and users to build interfaces through direct manipulation of graphical elements rather than programming them. 
% Additionally, Kurtenbach and Buxton \cite{kurtenbach1991issues} extended these concepts by integrating marking interactions in their GEdit system, showing that combining marking techniques with direct manipulation allows users to perform complex commands with simple gestures.  

While direct manipulation relies on interfaces and immediate feedback, natural language interaction (NLI) enables more conversational interactions \cite{liddy2001natural}.
Voice-activated commands and chatbots allow users to perform tasks through natural language, improving accessibility and reducing the cognitive load of traditional graphical user interfaces (GUIs) \cite{schmidhuber2021cognitive}.
By combining direct manipulation and natural language processing, Cohen et al. \cite{cohen1992role} developed the Shoptalk system, a prototype for information and decision making.
It allowed users to interact with databases and simulations using language alongside graphical actions like pointing and menus, enabling object description, temporal reasoning, and large-scale operations with immediate visual feedback.

% The system addressed the limitations of each modality, offering a more effective interface than using either alone.

\subsection{Natural Language Interaction}

While we mentioned different ways of editing network data and the use of NLP, the next step is to analyze how these natural language interactions occur within systems.
% Natural Language Interaction (NLI) involves communicating with systems and technologies using human language in a natural way \cite{ogden1997using}. 
% Familiar examples of NLI \cite{ogden1997using} include natural voice interactions with home virtual assistants such as Amazon Alexa.
One of the main benefits of NLI is that it enables individuals to engage with machines using conversational language, eliminating the need to learn commands or interfaces \cite{ogden1997using}. 
% NLI addresses issues of usability and accessibility, especially for individuals who may not be tech-savvy, by providing a means to interact with complex systems through simple, conversational language.
By taking advantage of NLI technology, researchers examined data visualization interaction models based on natural language \cite{shen2022towards,yu2019flowsense,wang2023enabling}.
For example, Yu et al. \cite{yu2019flowsense} developed a natural language interface called FlowSense, enabling users to construct and modify dataflow diagrams using natural language commands. 
% Through user studies, they found that users, including those with little prior experience, could efficiently create and manipulate complex visualizations through simple language inputs. 
% Similarly, Wang et al.~\cite{wang2023enabling} explored how large language models (LLMs) can be adapted to enable natural language interactions with mobile user interfaces (UIs).
% By developing and testing specific prompting techniques, the authors found that LLMs could perform these tasks with competitive accuracy, demonstrating the feasibility of using LLMs to facilitate natural language interactions on mobile UIs. 

While studies such as those by Wang et al. ~\cite{wang2021putting} and Power et al. ~\cite{power1998you} have demonstrated the overall efficiency of NLP in various domains, it is essential to acknowledge that NLP techniques are not without limitations and challenges.
For instance, NLP models often struggle with context ambiguity since the same phrase can have multiple meanings.
In particular, when dealing with complex network graph data, these shortcomings can become even more obvious due to the inherently interconnected nature of graphs.

To contribute to the existing body of literature on visualization and interaction techniques for networked data with a focus on node-link diagrams, we integrated and compared direct manipulation of graphs through click-based details on demand with more advanced methods like natural language interaction by utilizing LLMs to study the trade-offs and efficacy of each method. 
By comparing these forms of natural language interaction with traditional GUI-based methods, we can better assess which approach is more effective for different scenarios, such as making minor corrections or managing multiple data entries. 
% Existing research has not studied how natural language interfaces compare to traditional methods, nor has it explored the trade-offs involved in using NLP for editing and managing these node-link diagrams in different scenarios.
% For example, while natural language interaction techniques are expected to make it easier to make changes quickly, users may give their inputs in various forms, such as command-based instructions like ``delete node X'' or more descriptive language like ``There is no node X in there''. 
% Understanding these distinctions allows us to optimize the use of each interaction method based on the specific context and task requirements. 
% By comparing these forms of natural language interaction with traditional GUI-based methods, we can better assess which approach is more effective for different scenarios, such as making minor corrections or managing multiple data entries. 

\section{Experiment}
% \subsection{Visualizing Network data}
% \TODO{Talk about the need for vis? Preattentive processing? Change blindness?}

\begin{figure*}[t]
\centering
\includegraphics[width=1.8\columnwidth]{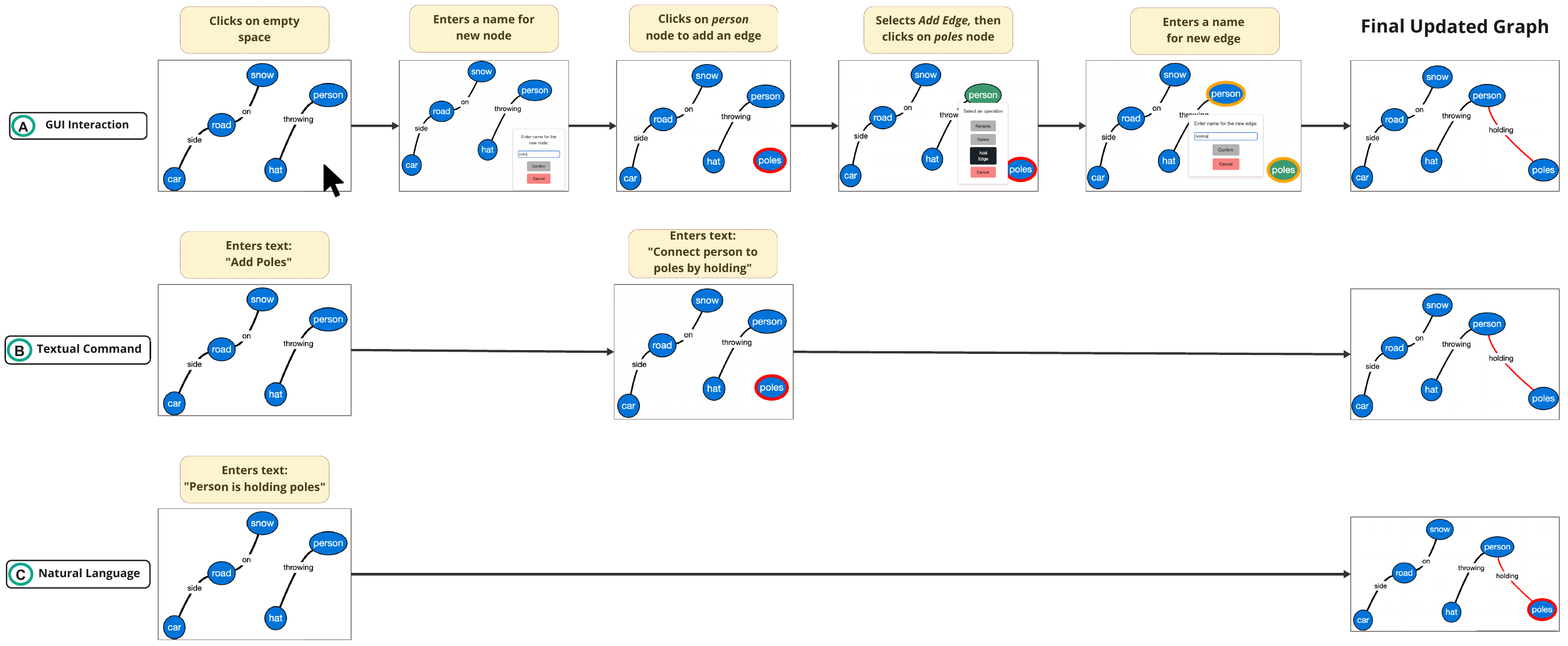} 
\caption{Screenshots of the interface before and after user commands: 
(A) GUI Interaction via clicking and menus, 
(B) Textual Command using structured input, 
(C) Natural Language using free-form input.}

\label{fig:all}

\end{figure*}

\subsection{Research Goals}
% The literature indicates that visualization and interaction methods significantly influence the process of gathering and updating knowledge in graph visualizations \TODO{Cite}. 
% \eric{need to come back and revise this after the related work. it will be easier to explain the research goals after introducing the foundations in the related work. the existing explanations in this section can be improved}

% \TODO{Link hypothesis and RQ to the conditions and measured}

Existing literature has demonstrated the efficiency of natural language interaction across various fields~\cite{yu2019flowsense,wang2023enabling}, highlighting its potential to simplify complex tasks. 
Building on this foundation, our research aims to explore how natural language techniques can enhance interaction within node-link or graph structures by comparing them to interaction through common GUI designs involving cursor selection, menus, and individually editing items. 
The specific research goals are as follows:

\begin{itemize}
\item \textbf{RQ1}: How does natural language input compare to GUI interaction for graph editing?
\item \textbf{RQ2}: How does method performance vary with graph size and extent of changes?
\item \textbf{RQ3}: How do command-style (e.g., ``rename X to Y'') vs. conversational text inputs (e.g., ``it's not an X, it's a Y'') affect editing?
\end{itemize}

To form our hypotheses, we considered the strengths and limitations of each interaction technique.
We expected natural language to outperform GUI methods due to its lower learning curve and ability to handle multiple edits in a single input, making it efficient for complex or contextual updates. 
However, we also predicted that GUI interaction might be more effective when only minimal edits are needed, as it allows for direct, precise adjustments without requiring detailed instructions.

Furthermore, we recognize that natural language can be used in multiple ways. Users might either describe a scene more narratively (e.g., ``a person is holding a bottle'') or issue concise commands (e.g., ``add a person and a bottle node'').
We expected that when users encounter minor issues in the graph, they will likely prefer command-like natural language inputs to fix specific elements (e.g., renaming a node or adjusting its properties) rather than describing the entire scene again.
This approach enables quicker, more targeted corrections, combining the flexibility of natural language with the precision of command-based interactions.
% Understanding the limitations of each method will help us leverage and apply them in the most appropriate contexts.

\begin{figure}[htb]
\centering
\includegraphics[width=0.9\columnwidth]{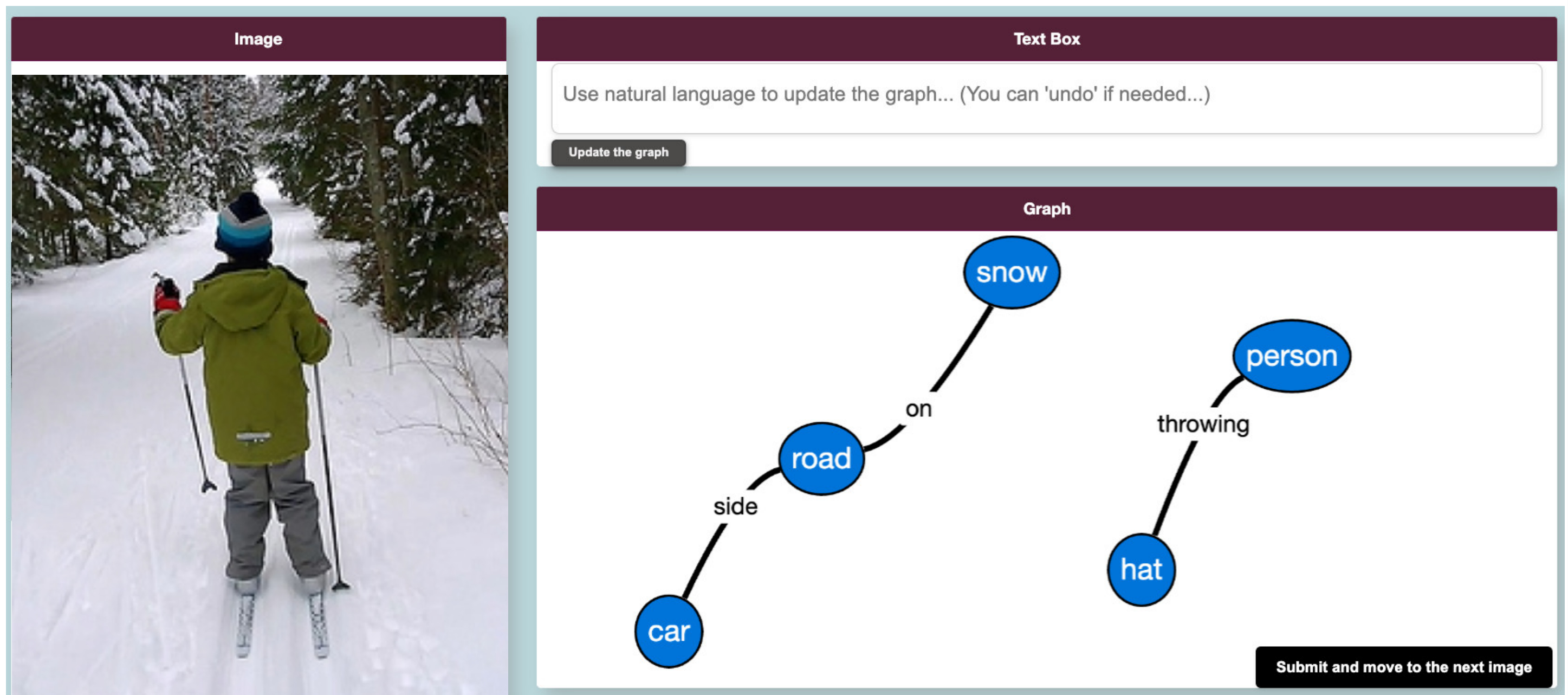} 
\caption{Screenshots of the study interface, including the text box for the {\nl} method. 
In each trial, participants updated the graph to match the image on the left. }
\label{fig:interface}
\end{figure}

% \eric{need to explain rationale for choice of task first, how it suits study goals and practical value. explain use of knowledge graphs. }
\subsection{Graph Editing Task}

To address our research goals through a user-study experiment, we needed a graph-editing task that involved various types of graph modifications at both the node and edge levels.
Knowledge graphs were chosen because they offer structured representations of graph data, where entities (nodes) and their relationships (edges) are clearly labeled and interconnected.
These graphs are further enhanced with features like metadata labels, hierarchical structures, and contextual information, making them ideal for modeling complex relationships and data attributes.
Knowledge can capture rich semantic relationships that apply to a wide range of real-world scenarios~\cite{theocharidis2009network,heist2020knowledge,orlandi2018leveraging}.
% They are commonly used in applied data systems, and typical forms of interaction---such as editing, expanding, and correcting values---ensures the reliability of the underlying data and correct functionality of any supported applications.
Interaction with knowledge graphs for data review also enables users to address errors, fill in missing information, and better connect related data, enhancing scalability and supporting the addition of new features~\cite{paulheim2017knowledge}.

As the basis for graph editing in the experiment, the experimental design required a way to encourage participants to make changes to a graph.
However, we also sought to avoid textual instructions for changes to prevent influencing the participants' use of natural language input by adapting the given text from instructions.
Thus, we designed the experimental task for graph editing based on an image prompt (Figure \ref{fig:vis}).
To this end, we utilized the \textit{GQA Dataset}~\cite{hudson2019gqa}, a highly regarded resource for visual reasoning that includes real-world images alongside their corresponding knowledge graphs that represent the objects and their spatial relationships.
Figure~\ref{fig:interface} shows an example from this dataset within the study interface.
% In the study, participants were tasked with updating (or creating) a knowledge graph describing the scene in the image.
Participants reviewed each image and its graph, then updated the graph by adding or correcting connections based on the image content.
While they had flexibility in the level of detail, the image simplicity encouraged consistent editing around primary objects. 
% Edits included adding nodes, creating edges, or modifying existing elements, using one of three methods described next.
 
% For each trial, participants viewed one image alongside a corresponding knowledge graph through an interactive web application. 
% Participants were assigned to one of the interaction types that we will describe in the next section to modify the existing graph or add new information to it based on the image.
% This task is closely aligned with real-world use cases and our study goal of keeping these data structures up-to-date.

% To account for different types of graph edits, including cases like smaller graphs with numerous inaccuracies or large graphs with only minor labeling errors, 
% Participants were asked to interact with a series of 24 images using a web-based interactive system.

% Participants could use graphical user interfaces (GUIs) to interact directly with the graph, clicking on nodes and edges to rename, delete, or add new edges as needed.
% Alternatively, they could use natural language commands to describe what they saw in the image or issue specific instructions, such as renaming a node or edge or adding an edge between two nodes.

% \begin{figure}[!t]
%     \centering
%     \includegraphics[width=0.45\textwidth]{figures/Teaser.pdf}
%     \caption{The experiment compared three graph editing methods: (A) GUI; (B) Textual Command; (C) Natural Language;}
%         \label{fig:teaser}
% \end{figure}

\subsection{Interaction Methods}

% Our research investigates alternative methods for interacting with and editing visual graph representations, aiming to overcome the limitations of traditional GUI-based interactions.
% Specifically, we examine how leveraging natural language processing (NLP) techniques can enhance the efficiency of graph interactions.

The experiment compared three different interaction methods for graph editing: \textit{\dm}, \textit{\nl}, and \textit{\nlc}.
For all versions, the study used the same visual design in the study application for displaying the graphs (Figure~\ref{fig:interface}).
To provide feedback to users after each input and help users easily see changes, the most recent modification to the graph---such as adding nodes or edges---were visually highlighted in red.
% This visual feedback was consistently available regardless of the interaction method used (Figure \ref{fig:all}).
However, the GUI condition provided additional sub-task feedback (such as context pop-up menus or node highlights to show selected nodes during node-linking actions)  to aid operations requiring multiple selections (Illustrated in Figure \ref{fig:all}).
The following subsections describe the three interaction methods and explain their differences.
 
\subsubsection{\dm}
% \TODO{Direct manipulation in HCI involves interacting directly with digital objects through actions like clicking, dragging, and dropping, } which provides immediate visual feedback and a sense of control \cite{shneiderman1983direct}. 
We implemented the {\dm} interaction method (shown in Figure \ref{fig:all}) as a standard method for interacting with a graph by cursor selection and contextual pop-up menus to select operations (i.e., add, remove, or rename) on the selected nodes or edges. 
% The method allows users to add, remove, and rename nodes and edges. 
% This method involves a pop-up menu that appears when a user clicks on a node or edge, enabling them to select the desired operation.
Users could also click on the white space of the box to create and name a new node.
Keyboard input was used for naming or editing the textual labels of nodes or edges.
For this method, the interface also included a ``Remove all'' button to delete all current nodes and edges in the graph.
Additionally, an ``Undo'' button was provided to revert the last action, whether it involved adding, renaming, or deleting a node or edge.

\subsubsection{\nlc}
\label{sec:nlc}

% While GUI Interaction emphasizes interfaces and immediate feedback, natural language processing (NLP) facilitates more conversational interactions \cite{liddy2001natural}. 
% To leverage this, we developed a technique that enables users to modify graphs using text input to enter commands for a wide range of graph operations.
% % as shown in Table \ref{tab:graph_operations}. 
% For instance, users can issue commands like ``Rename X to Y'' or ``Change X to Y'' or similar human-understandable variations (as shown in Figure \ref{fig:all}) to rename a node or edge. Additionally, commands such as ``Connect X to Y by Z'' or ``add a Z edge between X and Y'' or similar commands that allow users to add an edge Z between nodes X and Y. 
% However, users are not able to describe the scene, as that functionality is covered by our last method discussed in Section below (\ref{sec:nl}).
% This approach is designed for structured command inputs, utilizing predefined functions to directly add, remove, rename nodes, and update edges by filtering out scenarios with scene descriptions in the commands.

While GUI interaction relies on interfaces and immediate feedback, natural language processing (NLP) enables more conversational input \cite{liddy2001natural}. We developed a technique that lets users modify graphs through structured text commands for various operations. For example, users can enter commands like ``Rename X to Y'' or ``Connect X to Y by Z'' to rename nodes or add edges (Figure~\ref{fig:all}). This method is limited to structured edits and excludes scene descriptions, which are handled by the natural language approach in Section~\ref{sec:nl}. Commands are parsed to apply predefined functions for adding, removing, or renaming graph elements.

% Moreover, we leveraged ChatGPT API \cite{openai2024chatgpt} models to implement these features.
% Detailed instructions are then sent to the ChatGPT API \cite{openai2024chatgpt}, explaining how to update the graph based on the user's input.
% To improve the API model's understanding and accuracy, our system employs a technique known as few-shot prompting~\cite{brown2020language}.
% This involves providing the API with several examples of desired outputs along with the instructions.
% The system retrieves the relevant part of the dataset and identifies the specific graph corresponding to the image.
% The API processes this input and returns the updated graph data.
% The system parses this response and integrates the changes into the original dataset.
% By storing the current state of the dataset before any changes are made, the system ensures that users can revert to a previous state if necessary.
% \TODO{1. What models and prompts? 2. Validation of model outputs? 3. Response time taken to process?}\reza{DONE}

To study the potential of this approach, we leveraged the ChatGPT API \cite{openai2024chatgpt}, specifically the ``gpt-3.5-turbo'' model, to implement these features. 
Detailed instructions are then sent to the ChatGPT API \cite{openai2024chatgpt}, explaining how to update the graph based on the user's input.
These prompts break down all possible operations (e.g. add, rename, delete, etc.) and include examples for each through few-shot prompting~\cite{brown2020language}, thereby enhancing the model's performance and accuracy in understanding and executing user commands.

To validate the model output before running the study, we conducted multiple pilot studies and sessions with researchers who tested the system using challenging and unconventional commands.
Through several iterations, the model demonstrated high accuracy by effectively handling nearly all complex cases and natural language inputs. 
Importantly, the analysis of the data logged for Natural language conditions (51 out of 76 participants) showed that only a small number of trials (2.8\%) had to issue \textit{undo} commands to revert the changes, indicating the system's reliability and consistency.

In terms of performance metrics, the average processing time for both natural language and textual commands was 2.43 seconds.
This rapid response time ensures a near-seamless user experience, minimizing wait times to prevent potential user frustration or disengagement. 
As described in \textit{data processing} Section \ref{sec:measures}, we subtracted the processing time for each trial to allow a solid comparison without biasing the results, ensuring that performance metrics are directly tied to the specific model's capabilities.

% The system retrieves the relevant part of the dataset and identifies the specific graph corresponding to the image.
% The API processes this input and returns the updated graph data, which the system then parses and integrates into the original dataset.
% By storing the current state of the dataset before any changes are made, the system ensures that users can revert to a previous state if necessary.

% \eric{readers might want a more complete list of possible commands or types of operations. here, it is not even clear which types of actions or edits are supported. maybe a table would be useful?}

% \begin{figure*}[t]
% \centering
% \includegraphics[width=2\columnwidth]{figures/NLcommand.pdf} 
% \caption{Screenshots from the interface before and after running user command. 
% Updating graph using Textual Command method with the user command.}

% \label{fig:nlc}

% \end{figure*}

% \begin{figure}[t]
% \centering
% \includegraphics[width=1\columnwidth]{figures/NL.pdf} 
% \caption{Screenshots from the interface before and after running user command. 
% Updating graph using Natural Language method with the user command.}
% \label{fig:nl}
% \end{figure}

\subsubsection{\nl}
\label{sec:nl}

This method is an expanded version of the previously described {\nlc} technique from Section above (\ref{sec:nlc}). 
The {\nl} technique not only supports all the command types, but it also allows users to describe entire scenes using either a single command or multiple separate commands. 
For example, users can issue commands like ``person is holding poles'', ``person is wearing a hat'', and ``person has a green jacket'' either sequentially or combined in a single statement, and the graph will be updated accordingly (as shown in Figure \ref{fig:all}). 
This approach can accurately process any type of human-understandable text input to reflect changes in the graph. 
To implement these features, we leveraged models from the OpenAI API \cite{openai2024chatgpt} (Section \ref{sec:nlc}).

\subsection{Experimental Design}
% \TODO{Ordering for within subjects? duration of the study sessions and the chosen study apparatus? Differently-sized screens?} \reza{DONE}

In evaluating the different techniques for updating knowledge graphs, we also aimed to assess their efficiency across different scenarios and take into account variations in graph size and setup.
By comparing the number of changes made per unit of time or the number of operations required to implement a number of changes, we provide findings regarding the effectiveness of different methods for updating graphs as well as their trade-offs.
Moreover, participants reviewed an image alongside its corresponding knowledge graph (Figure~\ref{fig:interface}) and were tasked with adding or correcting information based on their assigned experimental condition.
% The method for interaction varied across participants, where each participant completed the task using the same method throughout the study so we could analyze participants' interaction details. 

To further analyze the impact of each interaction method on the graph, we divided the trials into two size groups for each participant, which we refer to as \textit{small} and \textit{large}. 
This division allows us to observe the effects of interaction methods on graphs of different complexities, helping to identify if certain methods are more effective based on the graph size. 
Graphs in the \textit{small} group had 4 or 5 nodes and 3 or 4 edges, while graphs in the \textit{large} group had 7 or 8 nodes and 6 or 7 edges. 

Additionally, to evaluate the effectiveness of these methods in scenarios with varying numbers of inaccuracies, we categorized the trials into three graph setup: \textit{Major} (50-60\% inaccuracies), where the task involves correcting a significant number of inaccuracies; \textit{Minor} (20-30\% inaccuracies), where the task involves addressing a relatively small number of inaccuracies; and \textit{Empty}, where the task requires starting from scratch and adding substantial new data to an initially empty graph.
It is important to note that we manually randomized all within-subject trials to minimize potential random effects and ensure a balanced distribution of conditions. 
This setup allows us to assess how different interaction methods perform in diverse contexts, from minor corrections to building a graph from scratch.
Therefore, the experiment followed a 3x2x3 mixed design with three independent variables: 1) \textit{Interaction Method} (between subjects) and 2) \textit{Graph Size} (within subjects), and 3) \textit{Graph Setup} (within subjects) as shown in Table \ref{tab:IV}.

\begin{table}[ht]
\caption{Independent variables and their levels for the 3×2×3 mixed-design user study}
\centering
\footnotesize
\setlength{\tabcolsep}{1pt}
\renewcommand{\arraystretch}{0.82}
\begin{tabular}{@{}p{0.3\columnwidth}|p{0.6\columnwidth}@{}}
\toprule
\multicolumn{1}{@{}p{0.3\columnwidth}@{}}{\centering\textbf{Independent Variables}} &
\multicolumn{1}{p{0.6\columnwidth}@{}}{\centering\textbf{Levels}} \\
\midrule
\multirow{3}{*}{\parbox[t]{0.3\columnwidth}{\centering Interaction Method\\\textit{(between-subjects)}}} &
1.~\dm \\ 
& 2.~\nlc \\
& 3.~\nl \\ \midrule
\multirow{2}{*}{\parbox[t]{0.3\columnwidth}{\centering Graph Size\\\textit{(within-subjects)}}} &
1.~Small (4--5 nodes \& 3--4 edges), 12 trials \\
& 2.~Large (7--8 nodes \& 6--7 edges), 12 trials \\ \midrule
\multirow{3}{*}{\parbox[t]{0.3\columnwidth}{\centering Graph Setup\\\textit{(within-subjects)}}} &
1.~Major (50--60\% inaccuracies), 8 trials \\
& 2.~Minor (20--30\% inaccuracies), 8 trials \\
& 3.~Empty (start with an empty graph), 8 trials \\ \bottomrule
\end{tabular}
\label{tab:IV}
\end{table}

\begin{figure*}[h]
\centering
\begin{minipage}[t]{0.46\textwidth}  % Adjust width to be equal
    \centering
    \includegraphics[width=\textwidth, height=0.88\textwidth]{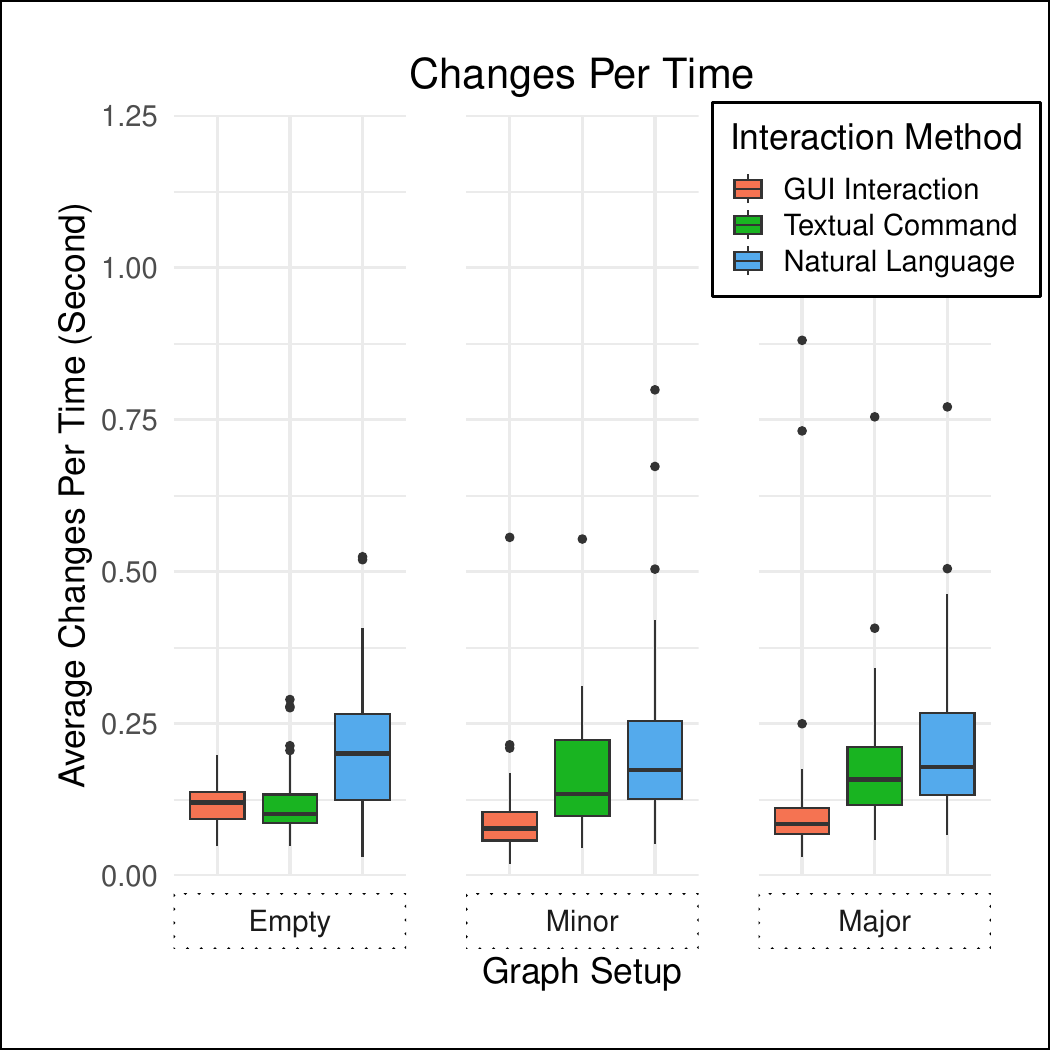}  % Set both width and height
    \caption{Average changes per time. Natural Language is significantly faster than all methods, and Textual Command outperforms GUI Interaction except in empty graph.}
    \label{fig:changesPerTime}
\end{minipage}%
\hspace{0.01\textwidth}  % Adjust spacing between figures
\begin{minipage}[t]{0.44\textwidth}  % Same width as the first minipage
    \centering
    \includegraphics[width=\textwidth, height=0.92\textwidth]{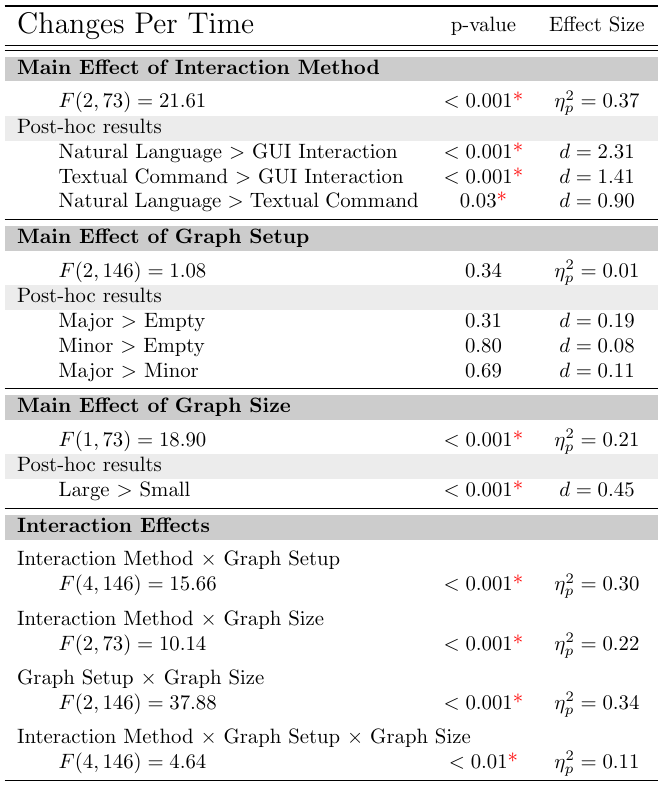}  % Set both width and height to match
    \caption{ANOVA and Posthoc Tukey HSD test results for Interaction Method differences on changes per time (* indicates a statistically significant difference at $p<0.05$)}
    \label{table:changesPerTime}
\end{minipage}
\end{figure*}

\subsection{Procedure and Participants}
The study was conducted online using a web application, allowing participants to work at their own pace without researcher intervention. Participation was voluntary, with extra credit offered. The study was approved by the Institutional Review Board (IRB).
Participants first provided informed consent and completed a demographic questionnaire (age, gender, education). They were then randomly assigned to one of three experimental conditions (Table~\ref{tab:IV}). After a tutorial and example trials, participants completed 24 image-graph tasks, with unlimited time to finish. They were required to make at least one change per graph to proceed. Interaction logs were collected anonymously, and data quality was ensured by excluding outliers and those who made few changes despite taking excessive time.
Eighty-nine participants took part, but after two rounds of quality checks (Section~\ref{sec:dataprocessing}), 76 were included in the analysis: \dm{} (25), \nlc{} (25), and \nl{} (26). All participants were students in computer science or human-centered computing courses, aged 18–40 (median 20). The final group included 45 males, 29 females, and 2 non-binary individuals.

\subsection{Measures}
\label{sec:measures}

During the online experiment, we logged participants' interactions to evaluate each interaction method.
For each image reviewed, we recorded 1) \textit{user response time}, 2) \textit{final version of the graph}, and 3) \textit{graph operation details}.
% User completion time includes \textit{response time} and \textit{action time}. 
% \textit{Response time} starts the moment users first see an image and ends when they submit the graph for that image.
% To eliminate potential noise data from the time spent thinking and reviewing each image and its associated graph, we also logged \textit{action time}.
% For direct manipulation, \textit{action time} begins with the first click.
% For both natural language methods, \textit{action time} starts with the first character typed in the text box.
% In all cases, \textit{action time} ends when users submit their final graph \reza{Reza: Do we need to explain action time while we are not using it for the analysis?} \eric{if we're not reporting any results or analysis for a measure, then don't report the measure}.
In addition to the final edited version of the graph submitted by each participant, we also logged all the \textit{graph operation details} for each action. 
For the direct manipulation method, this includes \textit{operations' name} such as adding, renaming, deleting, or removing nodes or edges.
For the natural language methods, it includes the \textit{user commands or text} used to edit the graph, along with the \textit{processing time} for each operation, which indicates the time taken on the backend to process the results.

\subsubsection{Data Processing}
\label{sec:dataprocessing}
To fairly compare interaction methods, particularly the time taken to make graph changes, we subtracted backend processing time from the response time in natural language conditions. This adjustment accounts for system delays unrelated to user performance and ensures comparability with direct manipulation, which has no backend delay. Backend time was excluded because it varies by model complexity.
Also, we applied two quality control steps: (1) participants with below 80\% accuracy across trials 1, 8, 16, and 24 were excluded to ensure consistent engagement (8 excluded); (2) trials with response time Z-scores above 2 were removed (4.2\% of trials), and participants with more than three such trials were also excluded (5 more excluded), resulting in 13 total exclusions from the 89 participants.

\subsection{Metrics}
After logging and filtering the data, we calculate the number of changes made to each graph. This helps assess the speed and quality of edits across interaction methods and graph types. We define two metrics based on these changes to present our results.

% \subsubsection{Graph Changes Heuristic}
% We detect differences between the initial graph ($G_1$) and final graph ($G_2$) by identifying changes in nodes and edges: additions, deletions, and renamings.
% The heuristic works in two phases.  
% For nodes, matches with shared edges are marked as renamed; unmatched nodes are added or removed.  
% For edges, existing edges in both graphs are kept; missing ones are marked as renamed or removed, and new edges as added.  
% Total changes are computed by summing renamed, added, and removed nodes and edges.
% These metrics address our research questions by quantifying user performance in terms of edit speed and precision across interaction methods, input formats, and graph complexity.

\subsubsection{Graph Changes Per Time}
For each image reviewed, we captured 1) \textit{User completion time}, 2) \textit{Final version of the graph}, and 3) \textit{Operation Details} as explained in Section \ref{sec:measures}.
Then, we detected differences between the initial graph and final graph to determine the number of changes each participant made for each image. The \textit{graph changes per time} metric was calculated by dividing the \textit{number of changes} by the \textit{response time} for each trial or image.
This metric allows us to compare the advantages and trade-offs of each interaction method by quantifying how quickly and effectively users can modify the graph.

\subsubsection{Graph Changes Per Action}
We used the \textit{graph changes heuristic} to determine the number of changes each participant made for each image by comparing the initial and final graphs. The \textit{graph changes per action} metric was calculated by dividing the \textit{number of changes} by the \textit{number of actions} for each trial or image. This metric enables us to compare the advantages and trade-offs each interaction method by determining whether one method allows users to achieve their goals with fewer or more precise actions.

\section{Results}

The experiment evaluated graph manipulation efficiency across interaction methods, graph sizes, and setups. As the data violated normality and homogeneity assumptions, we used ARTool \cite{wobbrock2011aligned} for nonparametric factorial analysis via aligned rank transformation. A three-way mixed ANOVA tested the effects of interaction method, graph size, and setup. For significant main effects, we performed post-hoc paired t-tests with Tukey correction using ARTool \cite{elkin2021aligned}, and report results with partial eta squared (\(\eta^2_p\)) as the effect size.

\subsection{Changes Per Time}

The study's results highlight findings regarding how different interaction methods for editing graphs impact the average changes per time for the varied sizes and expected amount of changes in each graph (Figure~\ref{fig:changesPerTime}, Table~\ref{table:changesPerTime}).
% The results for \textit{changes per time} are shown graphically in Figure~\ref{fig:changesPerTime}, and the analysis results are summarized in Figure~\ref{table:changesPerTime}.
The test detected a significant main effect of the \textit{interaction method} on \textit{changes per time}.
Post-hoc tests revealed that {\nl} resulted in significantly higher changes per time compared to both {\dm} ($p < 0.001$, Cohen's $d = 2.31$) and {\nlc} ($p < 0.05$, Cohen's $d = 0.9$), indicating it is a more effective interaction method for making faster changes. 
Also, {\nlc} had significant higher \textit{changes per time} than {\dm} ($p < 0.001$, Cohen's $d = 1.41$).

Additionally, the results also show a significant interaction effect between \textit{interaction method} and \textit{graph setup} (Figure~\ref{table:changesPerTime}), indicating that the impact of interaction methods on performance varies with the type of graph setup.
For example, in the case of an empty \textit{graph setup}, the test did not detect a significant interaction effect between the GUI Interaction and Textual Command.
However, in the case of minor and major \textit{graph setup}, the interaction effect was significant.
Under these conditions, the Textual Command method significantly outperformed GUI Interaction, showing an advantage for relying on text commands when addressing major graph inaccuracies. 
Specifically, for both the small \textit{graph size} ($p < 0.001$, $d = 1.46$) and the large \textit{graph size} ($p < 0.01$, $d = 1.86$) in the major inaccuracies \textit{graph setup}, Textual Command yielded significantly higher \textit{changes per time}.

Although no significant difference was found between Natural Language and Textual Command in the minor and major \textit{graph setup} with large \textit{graph size}, a significant difference emerged for the empty \textit{graph setup}, where Natural Language showed higher changes per time ($p < 0.01$, $d = 1.63$).

This interaction effect highlights the advantages of each method depending on the nature of the task.
GUI Interaction proves more efficient for adding new information to the graph, as it allows for quicker visual adjustments without the need for multiple text commands, which can be cumbersome when dealing with repetitive inputs. 
However, for correcting major inaccuracies or modifying existing details, Textual Commands enable faster edits by letting users describe changes directly, avoiding manual selection, pop-up menus, and switching between keyboard and mouse.

% Furthermore, the graph size also played a significant role, both as a main effect and in its interaction with the interaction method. Larger graphs might tend to demand more cognitive resources, influencing how users interact with and modify them. The significant difference between large and small graph sizes in post-hoc tests ($p < 0.001$, Cohen's $d = 0.41$) shows the effect of graph size on the different numbers of changes.

% As shown in Figure~\ref{fig:changesPerTime} and Table~\ref{table:changesPerTime}, there is a significant main effect of the \textit{interaction method} on \textit{changes per time}. Post-hoc test found that {\nl} had significantly higher than {\dm} ($p < 0.001$, Cohen's d = 2.13) as well as {\nlc} ($p < 0.05$, Cohen's $d = 0.83$). Also, {\nlc} had significantly higher than {\dm} ($p < 0.001$, Cohen's d = 1.3)

\begin{figure*}[h]
\centering
\begin{minipage}[t]{0.46\textwidth}
    \centering
    \includegraphics[width=\textwidth, height=0.89\textwidth]{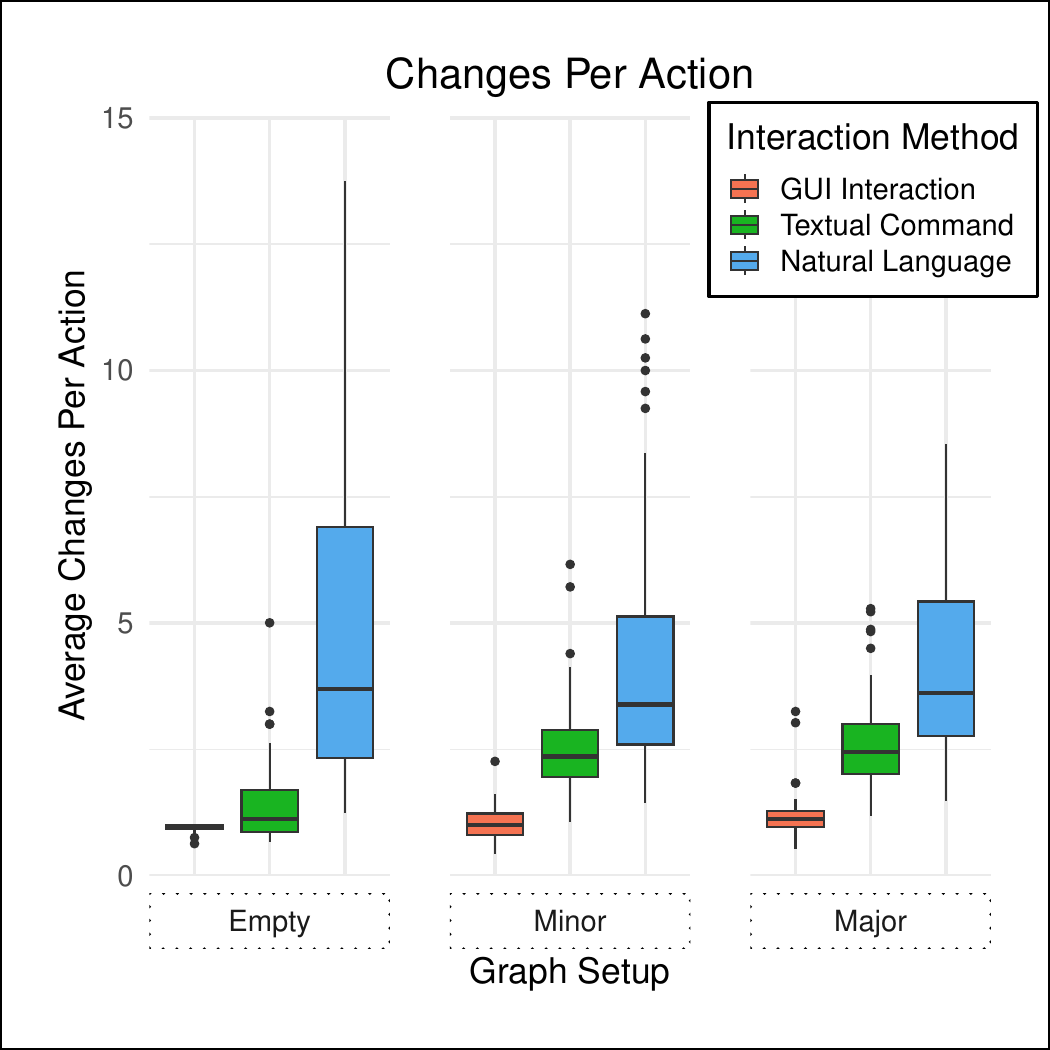}
    \caption{Average changes per action. Natural Language requires less effort, and Textual Command achieves more changes per action than GUI Interaction in all cases. Higher values reflect more graph changes per action, such as renaming in GUI or text inputs in other methods.}
        \label{fig:changesPerOperation}
\end{minipage}%
\hspace{0.01\textwidth}
\begin{minipage}[t]{0.43\textwidth}
    \centering
    \includegraphics[width=\textwidth, height=0.95\textwidth]{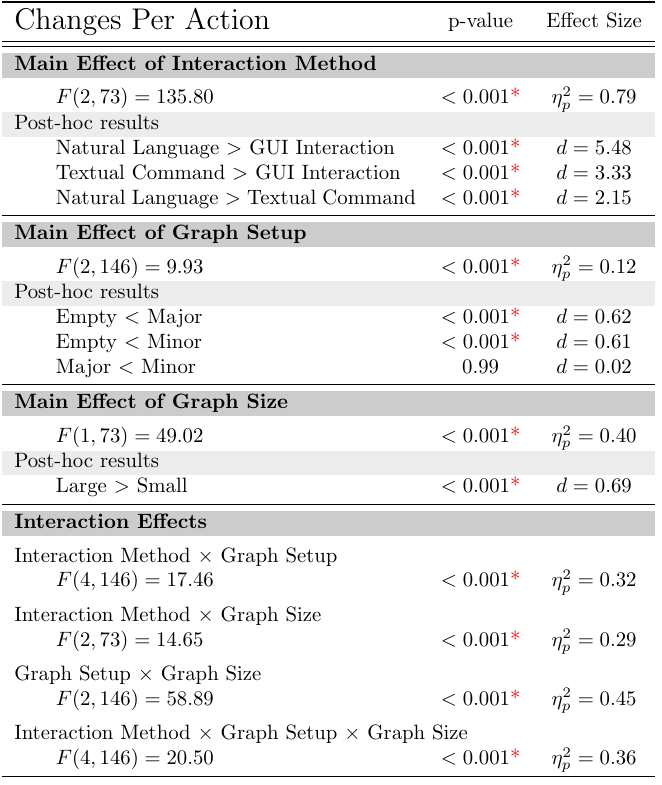}
    \caption{ANOVA and Tukey HSD results for interaction method differences in changes per action (* indicates $p<0.05$). GUI actions include each operation (e.g., add node/edge); for other methods, actions refer to each text input submitted.}
    \label{table:changesPerOperation}
\end{minipage}
\end{figure*}

\subsection{Changes Per Action}

The results for \textit{changes per action} (Figures~\ref{fig:changesPerOperation} and~\ref{table:changesPerOperation}) show a significant main effect of \textit{interaction method}, indicating clear differences in the number of changes achieved per action across methods.
Post-hoc tests revealed that {\nl} resulted in significantly higher changes per action compared to both {\dm} ($p < 0.001$, Cohen's $d = 5.48$) and {\nlc} ($p < 0.001$, Cohen's $d = 2.15$). 
This suggests that the {\nl} method enables users to accomplish more modifications per action, likely due to the ability to describe or command multiple changes at once rather than performing single, discrete actions as required in {\dm}.
Furthermore, the {\nlc} method also showed higher changes per action than the {\dm} method ($p < 0.001$, Cohen's $d = 3.33$). 
This indicates that {\nlc} allows users to input several commands to modify the graph, whereas {\dm} typically involves more granular, individual actions.

Additionally, the analysis for \textit{changes per action}   found a significant interaction effect between \textit{interaction method} and \textit{graph setup} ($F(4,365) = 23.38, \eta^2_p = 0.2$).
This effect indicates that the influence of interaction methods on performance varies depending on the type of graph setup.
This is similar as what described for \textit{changes per time} with the fact that it can be seen both for minor inaccuracies \textit{graph setup} ($p < 0.001$, $d = 2.99$) as well as major inaccuracies ($p < 0.001$, $d = 2.81$) \textit{graph setup} but only for large \textit{graph size}.

% \subsection{Completion Time}
% As shown in Figure~\ref{fig:completionTime} 

% \begin{figure}[h]
% \centering
% \includegraphics[width=0.6\columnwidth]{figures/AverageRT.pdf} 
% \caption{Average completion time}
% \label{fig:completionTime}
% \end{figure}

\section{Discussion}
% \TODO{Address RQs here and provide broader application}

\subsection{Interpretation of Results}
The study aimed to evaluate the trade-offs of different interaction methods with graph data across various graph complexity and graph setups.
The results revealed that the Natural Language method consistently outperformed other approaches, particularly in terms of speed and the efficiency of implementing changes over time. 
It is important to note that, although this was an online study, we conducted several iterations of quality checks and outlier removal, as detailed in Section \ref{sec:dataprocessing}, to ensure that the recorded time was logical and minimally affected by the inherent interruptions of online studies.
As shown in Figure \ref{fig:changesPerTime} and \ref{table:changesPerTime}, which present the average number of changes over time, the Natural Language method enabled significantly faster modifications compared to other methods. 
This was especially evident when the graph was empty, indicating scenarios where new information needed to be added.
Additionally, for graphs with minor and major inaccuracies, where the goal was to edit errors in addition to adding new information, the Natural Language method still achieved a higher rate of changes over time. 
However, the performance gap between Natural Language and other methods was narrower in these cases. 

Figure \ref{fig:changesPerOperation} and \ref{table:changesPerOperation}, which display the average number of changes per action, highlight the relative difficulty users encounter when modifying the graph by taking advantage of multiple actions such as clicking or submitting text entry.
The Natural Language and Textual Command particularly excelled in graphs with minor and major inaccuracies due to the efficiency of implementing multiple changes in each input.
For example, a single input can include multiple operations and many details, which is not possible with a single execution through GUI Interaction.
In contrast, for \textit{empty graphs}, where users begin from scratch, no significant difference was observed between GUI Interaction and Textual Command. 
Nonetheless, the Natural Language method consistently outperformed both GUI Interaction and Textual Command across all scenarios.

Notably, after analysis of user inputs for the Natural Language method based on a manual review of all inputs logged from participants, we found that an average of 51\% of inputs were text-based commands.
Participants, despite having the flexibility to use various types of text input, frequently opted for Textual Commands when tasks involved editing information.
Figures \ref{fig:changesPerTime} and \ref{fig:changesPerOperation} illustrate this pattern, showing that while Natural Language usage is significantly higher than Textual Commands in the \textit{empty} graph setup, the gap narrows in both the \textit{minor} and \textit{major} \textit{graph setup}, reinforcing the observation.
This supports use of text-based commands for editing purposes.
The preference for Textual Commands likely stems from allowing precise edits without re-describing the entire scene.
On the other hand, Natural Language allows them to describe their thoughts with the freedom of language they are comfortable with, resulting in an advantage when combined with textual commands.

\subsection{Natural Language Use Cases in Different Application Domains}
\label{sec:discuss-other-cases}

Considering the strong performance of natural language for editing knowledge graphs, this method has potential for broader applications.
Integrating natural language interaction into data visualization \cite{shen2022towards,wang2022towards} marks a shift from traditional GUI-based approaches.
While tools like~\cite{henry2007nodetrix,dunne2012graphtrail} enable flexible exploration, they often require navigating complex menus, switching visual forms, and mastering techniques like zooming or filtering.
Natural language offers a simplified alternative by letting users issue commands such as ``Show me a bar chart of sales data grouped by region'' or ``Highlight the most connected nodes.'' This reduces the effort needed to generate visualizations, making them more accessible to users.
Additionally, natural language supports features like interactive annotations, allowing users to label visual elements with contextual information \cite{srinivasan2018augmenting}. These annotations can also be interactive, enabling dynamic exploration and reinforcing feedback loops during data updates. This aligns with our research goal of streamlining data interaction through natural and intuitive modalities.

Natural language interfaces are especially valuable in HITL applications, where users provide feedback to improve model performance \cite{li2016dialogue, chen2024learning, endert2014human}. These interfaces offer a way for users to update data or provide input without navigating complex interfaces.
For example, users can request dataset updates or corrections as new information becomes available.
Since our study used graph data from GQA dataset \cite{hudson2019gqa}, commonly used in visual reasoning, it demonstrates how natural language interaction can support HITL tasks like data correction and annotation. 
Similar approaches can be extended to other data sources or AI-powered systems.

\section{Conclusion}

% This paper demonstrates the effectiveness of natural language interaction for manipulating node-link graph visualizations through text input. A user study shows that both free-form natural language and command-style textual methods serve as viable alternatives to traditional GUI-based editing, offering flexibility and accessibility for graph updates. The results provide strong evidence for the feasibility of natural language approaches and highlight the value of allowing descriptive input beyond rigid command formats.

% However, one limitation is that the observed efficiency in the Natural Language (NL) condition—measured by changes per time and per action—was influenced by design differences. 
% In the NL condition, users could combine node and edge creation in one input, while the GUI required sequential steps (Figure~\ref{fig:all}).
% This structural difference likely contributed to higher performance in the NL condition and should be considered when interpreting results.
% Additionally, the study was limited in the range of graph sizes and editing tasks due to the constraints of an online setting.
% Exploring a broader variety of graph complexities could provide deeper insights into the strengths and weaknesses of each interaction method. Although node-link diagrams are widely used, natural language techniques may require adaptation for alternative visual representations.

This paper demonstrates that natural language input offers an effective alternative to GUI-based graph editing, enabling flexible interaction with node-link visualizations. A user study confirms the benefits of supporting descriptive input.
However, one limitation is that NL efficiency may be inflated due to design differences, as users could perform combined actions in a single input, unlike the step-by-step process required in the GUI (Figure~\ref{fig:all}).
Also, the study was limited in the range of graph sizes and editing tasks due to the constraints of an online setting.
Future work should explore larger graphs and diverse visual representations in addition to node-link diagrams.

\begin{acks}
This was supported by the DARPA ECOLE Program HR00112390063.
\end{acks}

\appendix

\bibliographystyle{ACM-Reference-Format}
\bibliography{sample-base,software}

\end{document}